\documentclass[conference]{IEEEtran}
\usepackage{algorithm,algorithmic}
\makeatletter
\newcommand\fs@betterruled{%
  \def\@fs@cfont{\bfseries}\let\@fs@capt\floatc@ruled
  \def\@fs@pre{\vspace*{5pt}\hrule height.8pt depth0pt \kern2pt}%
  \def\@fs@post{\kern2pt\hrule\relax}%
  \def\@fs@mid{\kern2pt\hrule\kern2pt}%
  \let\@fs@iftopcapt\iftrue}
\floatstyle{betterruled}
\restylefloat{algorithm}
\makeatother
\newcommand{\squeezeup}{\vspace{-4mm}}

\newcommand{\squeezeupann}{\vspace{-1mm}}

\usepackage{cite}
\usepackage{amsmath,amssymb,amsfonts}
\usepackage{algorithmic}
\usepackage{graphicx}
\usepackage{textcomp}
\usepackage{comment}
\usepackage{enumitem}

\def\BibTeX{{\rm B\kern-.05em{\sc i\kern-.025em b}\kern-.08em
    T\kern-.1667em\lower.7ex\hbox{E}\kern-.125emX}}
\begin{document}

\title{Simultaneous Downlink Data Transmission and Uplink Channel Estimation with Reduced Complexity Full Duplex MIMO Radios}

\author{\IEEEauthorblockN{Md Atiqul Islam\IEEEauthorrefmark{2}, George C. Alexandropoulos\IEEEauthorrefmark{4}, and Besma Smida\IEEEauthorrefmark{2}}
\IEEEauthorblockA{{\IEEEauthorrefmark{2}Department of Electrical and Computer Engineering, University of Illinois at Chicago, USA}\\
\IEEEauthorrefmark{4}Department of Informatics and Telecommunications, National and Kapodistrian University of Athens, Greece\\
emails: \{mislam23,smida\}@uic.edu, alexandg@di.uoa.gr
}}
\maketitle

\begin{abstract}
In this paper, we study Full Duplex (FD) Multiple-Input Multiple-Output (MIMO) radios for simultaneous data communication and control information exchange. Capitalizing on a recently proposed FD MIMO architecture combining digital transmit and receive beamforming with reduced complexity multi-tap analog Self-Interference (SI) cancellation, we propose a novel transmission scheme exploiting channel reciprocity for joint downlink beamformed information data communication and uplink channel estimation through training data transmission. We adopt a general model for pilot-assisted channel estimation and present a unified optimization framework for all involved FD MIMO design parameters. Our representative Monte Carlo simulation results for an example algorithmic solution for the beamformers as well as for the analog and digital SI cancellation demonstrate that the proposed FD-based joint communication and control scheme provides $1.4\times$ the downlink rate of its half duplex counterpart. This performance improvement is achieved with $50\%$ reduction in the hardware complexity for the analog canceller than conventional FD MIMO architectures with fully connected analog cancellation.
\end{abstract}


\section{Introduction}
Full Duplex (FD) communication technology has the potential of doubling the spectral efficiency and simplifying the control information exchange over conventional frequency- and time-division duplexing systems through concurrent UpLink (UL) and DownLink (DL) communication in the same frequency and time resources \cite{sabharwal2014band,bharadia2013full,smida2017reflectfx,duarte2012experiment,islam2019comprehensive,chen2019wideband,askar2019polarimetric}. Exploitation of Multiple-Input Multiple-Output (MIMO) communication provides an efficient system performance boost due to increasing the system's spatial Degrees of Freedom (DoF) offered by the plurality of Transmitter (TX) and Receiver (RX) antennas \cite{riihonen2011mitigation,everett2016softnull,alexandropoulos2017joint,bharadia2014full,masmoudi2017channel,anttila2014modeling,iimori2019mimo}. Thus, incorporating FD with MIMO operation can contribute to the demanding throughput and latency requirements of fifth Generation (5G), and beyond, wireless communication systems with efficient utilization of the limited spectrum resources.\par

The simultaneous transmission and reception in FD systems equipped with a TX and a RX results in an in-band Self Interference (SI) signal at the reception side due to the limited TX and RX isolation. Therefore, a combination of propagation domain isolation, analog domain suppression, and digital SI cancellation techniques are adopted in practice to suppress the strong SI below the noise floor. In FD MIMO systems, the suppression techniques are particularly challenging due to higher SI components, as a consequence of the increased number of transceiver antennas. Propagation domain isolation is accomplished by pathloss or antenna directionality, whereas analog domain suppression in Single-Input Single-Output (SISO) systems is achieved by subtracting a processed copy of the TX signal from the RX inputs to avoid saturation of RX's Radio Frequency (RF) chain~\cite{sabharwal2014band}. Analog SI cancellation in FD MIMO systems can be implemented through SISO replication. However, the hardware requirements for such an approach scale with the number of TX/RX antennas, rendering the implementation of analog SI a core design bottleneck. In \cite{riihonen2011mitigation,everett2016softnull}, authors presented spatial suppression techniques that alleviate the need of analog SI cancellation relying solely on digital TX/RX beamforming. In \cite{alexandropoulos2017joint}, a joint design of multi-tap analog cancellation and TX/RX beamforming, where the number of taps does not scale with the product of TX and RX antenna elements, was proposed. According to this work, every analog tap refers to a line of fixed delay, variable phase shifter, and attenuator. The authors in \cite{islam2019unified} presented a unified beamforming approach including both Analog and Digital (A/D) SI cancellation that further improves the achievable rate performance under practical transceiver imperfections.
\begin{figure*}[!t]
\centering
\includegraphics[width=0.95\textwidth]{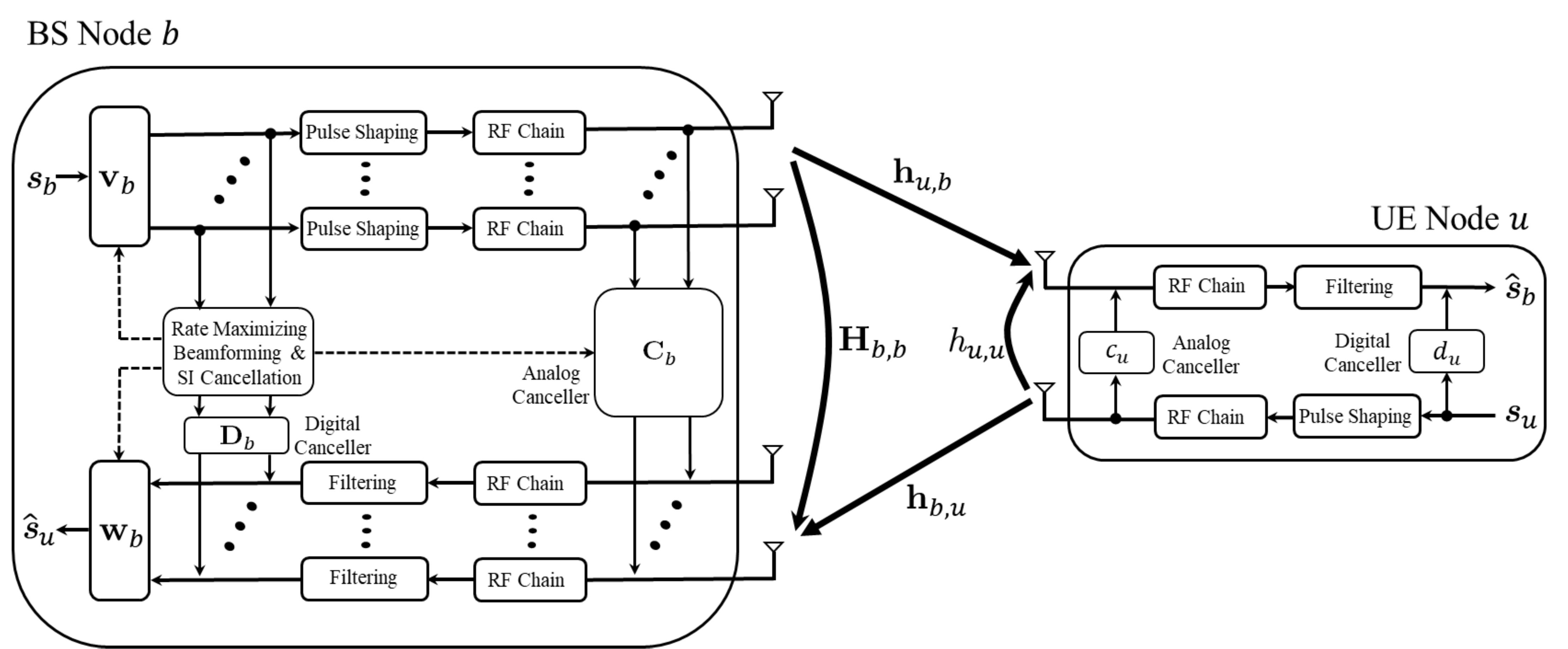}
\caption{The considered system model for simultaneous DL data transmission and UL pilot-assisted channel estimation. Both BS node $b$ and UE node $u$ deploy A/D SI cancellation with the former node $b$ realizing \cite{islam2019unified}'s reduced complexity multi-tap analog canceller.}
\squeezeup
\label{fig: transmission}
\end{figure*}

Recently, in \cite{du2015mu,du2016sequential,mirza2018performance}, simultaneous DL data transmission and UL Channel State Information (CSI) reception has been considered at a FD MIMO Base Station (BS) serving multiple Half Duplex (HD) User Equipment (UE) nodes. Specifically, those studies exploit the additional channel offered by the FD operation for receiving UL training symbols at BS from UEs to estimate the DL channels leveraging channel reciprocity, while at the same time transmitting the DL payload to UEs. In \cite{du2015mu}, the authors presented an adaptive beamforming approach for channel acquisition with DL data transmission, whereas in \cite{du2016sequential}, a sequential beamforming technique was proposed for multi-user MIMO systems improving the DL rate performance. In \cite{mirza2018performance}, an integrated FD-HD model is presented for concurrent DL transmission and UL CSI acquisition. However, the latter approaches are based on switching schemes between the FD and HD modes, which render the implementation of SI cancellation quite challenging. Furthermore, the adopted multi-user models consider perfect analog cancellation that is based on conventional FD MIMO architectures with fully connected analog cancellation interconnecting all TX antenna elements in the FD node with all its RX antennas.

In this paper, we capitalize on the FD MIMO transceiver architecture of \cite{islam2019unified} combining TX/RX beamforming with A/D SI cancellation, and present a point-to-point two-way communication system for simultaneous DL information data transmission and UL CSI estimation. Exploiting channel reciprocity and relying exclusively on FD operation, the proposed system performs joint digital TX beamforming for DL communication and UL pilot-assisted CSI estimation. Considering realistic modeling for imperfect channel estimation, we present a unified optimization framework for joint DL rate optimization and accurate CSI estimation. Our simulation results showcase superior achievable DL rate performance for the proposed FD-based transmission scheme compared to conventional HD systems with up to $50\%$ less analog cancellation hardware complexity than conventional FD MIMO architectures with fully connected analog cancellation.
 
\textit{Notation:} Vectors and matrices are denoted by boldface lowercase and boldface capital letters, respectively. The transpose, Hermitian transpose, and conjugate of $\mathbf{A}$ are denoted by $\mathbf{A}^{\rm T}$, $\mathbf{A}^{\rm H}$, and $\mathbf{A}^*$, respectively, and $\det(\mathbf{A})$ is $\mathbf{A}$'s determinant, while $\mathbf{0}_{m\times n}$ ($m\geq2$ and $n\geq1$) represents the $m\times n$ matrix with all zeros. $\|\mathbf{a}\|$ stands for the Euclidean norm of $\mathbf{a}$. $[\mathbf{A}]_{i,j}$, $[\mathbf{A}]_{(i,:)}$, and $[\mathbf{A}]_{(:,j)}$ represent $\mathbf{A}$'s $(i,j)$-th element, $i$-th row, and $j$-th column, respectively, while $[\mathbf{a}]_{i}$ denotes the $i$-th element of $\mathbf{a}$. $\mathbb{C}$ represents the complex number set, $\mathbb{E}\{\cdot\}$ is the expectation operator, and $|\cdot|$ denotes the amplitude of a complex number. $x\sim\mathcal{CN}(0,\sigma^2)$ represents a circularly symmetric complex Gaussian random variable with zero mean and variance $\sigma^2$.

\section{System and Signal Models}\label{sec: Sig1}
A a point-to-point two-way wireless communication system is considered with a FD MIMO BS node $b$ comprising $N_b$ TX and $N_b$ RX antennas and a FD single-antenna UE node $u$, as shown in Fig. \ref{fig: transmission}. Each antenna is attached to dedicated TX/RX RF chain at both nodes. A multi-tap analog SI canceller is applied in the FD BS node $b$, whereas a single-tap SI canceller is used for the UE node $u$. The FD BS node is capable of performing digital TX/RX beamforming realized for simplicity with linear filters. We consider UL/DL channel reciprocity and focus on joint data transmission and CSI estimation, where the DL is intended for information data communication while the UL is used for transmitting training signals from UE to BS. The UL CSI estimation is used for designing the DL precoder to digitally process the data signals before transmission.

We assume that BS node $b$ transmits the complex-valued information data symbol ${s}_b$ (chosen from a discrete modulation set) using the unit norm digital precoding vector $\mathbf{v}_b\in\mathbb{C}^{N_b\times 1}$. Similarly, the single-antenna UE node $u$ sends the training symbol ${s}_u$ through the UL channel. The signal transmissions at nodes $b$ and $u$ are power limited to ${P}_b$ and ${P}_u$, respectively. Specifically, the DL signal is such that $\mathbb{E}\{\|\mathbf{v}_b{s}_b\|^2\}\leq {P}_b$, whereas the UL signal is constrained as $\mathbb{E}\{|{s}_u|^2\}\leq {P}_u$. As both nodes are capable of FD operation, the simultaneously transmitted DL data and UL training signals induce SI in the RXs of BS and UE, respectively. We consider the Rician fading model for the SI channels denoted by $\mathbf{H}_{b,b} \in \mathbb{C}^{N_b\times N_b}$ for the BS and ${h}_{u,u}\in\mathbb{C}$ for the UE with Rician factor $K$ and pathlosses $l_{b,b}$ and $l_{u,u}$ at node $b$ and $u$, respectively \cite{duarte2012experiment}. The Rayleigh faded UL and DL channels, denoted respectively by  $\mathbf{h}_{b,u},\mathbf{h}_{u,b}^{\rm T}\in \mathbb{C}^{N_b\times 1}$, are modeled as Independent and Identically Distributed (IID) $\mathcal{C}\mathcal{N}(0,l_{b,u})$ and $\mathcal{C}\mathcal{N}(0,l_{u,b})$, where, $l_{b,u}$ and $l_{u,b}$ are the UL and DL pathlosses, respectively. 

At the FD BS node $b$, the induced SI is first suppressed at the inputs of its RX RF chains using multi-tap analog cancellation, as shown in Fig. \ref{fig: transmission}. For further cancellation of the residual SI, digital cancellation is also applied at the RX's baseband. After both cancellation approaches, the received baseband signal $\mathbf{y}_{b}\in \mathbb{C}^{N_b\times 1}$ at node $b$ is expressed as
\begin{equation}\label{Eq: sig1}
    \begin{split}
        \mathbf{y}_{b} \triangleq \left(\mathbf{H}_{b,b}+\mathbf{C}_{b}+\mathbf{D}_b\right)\mathbf{v}_b{s}_{b} + \mathbf{h}_{b,u}{s}_{u} + \mathbf{n}_{b}, 
    \end{split}
\end{equation}
where $\mathbf{n}_{b}\in \mathbb{C}^{N_b\times 1}$ is the zero-mean Additive White Gaussian Noise (AWGN) with variance $\sigma^2_{b}$. In the latter expression, $\mathbf{C}_{b}\in \mathbb{C}^{N_b\times N_b}$ and $\mathbf{D}_b\in \mathbb{C}^{N_b\times N_b}$ are the matrices representing A/D SI canceller at BS, respectively.

The analog SI cancellation at BS node $b$ is performed following the approach of \cite{alexandropoulos2017joint} that deploys $N$-tap analog cancellation. First, the DL information data signals in the outputs of the $N_b$ TX RF chains are fed to the analog canceller via $N$ MUltipleXers (MUXs). The outputs from the MUXs are attenuated and phase shifted to emulate the over-the-air SI channel $\mathbf{H}_{b,b}$, and connected to the inputs of the $N_b$ RX RF chains via $N$ DEMUltipleXers (DEMUXs). The cancellation matrix $\mathbf{C}_{b}$ captures the configuration of the MUXs/DEMUXs and the canceller tap values. For a given number $N$, there are different ways to connect them between the TX and RX antennas. Each of these configurations refers to a unique $\mathbf{C}_{b}$ realization, which corresponds to a specific placement of the $N$ tap values inside $\mathbf{C}_{b}$ and its remaining ${N_b}^2-N$ elements are set to zeros. Placing $N$ taps at the elements with the $N$ largest in amplitude elements of $\mathbf{H}_{b,b}$ can be a reasonable realization of the SI matrix $\mathbf{C}_{b}$. Other approaches can be orderly column-by-column or row-by-row placement of $N$ taps. For example, a $N$-tap analog cancellation matrix, where the tap values are placed in consecutive rows of the matrix, is given by
\begin{equation}\label{Eq: analog_canc}
    \begin{split}
        \mathbf{C}_{b}=\begin{bmatrix}-[{\mathbf{H}}_{b,b}]_{(1:\frac{N}{N_b},:)}\\
                        \mathbf{0}_{(\frac{N}{N_b}+1:N_b,:)}
                        \end{bmatrix}.
    \end{split}
\end{equation}

The residual SI signal after analog cancellation in the digital domain, which is modeled as $\left(\mathbf{H}_{b,b}+\mathbf{C}_{b}\right)\mathbf{v}_b{s}_{b}$ in \eqref{Eq: sig1}, is suppressed using a digital cancellation approach. As shown in \cite{islam2019unified}, digital cancellation suppresses both linear and non-linear SI induced by the TX RF chains. However, in this paper, we consider only linear SI for brevity. Matrix $\mathbf{D}_b$ represents the digital canceller, which is multiplied with the precoded DL signals and added to the outputs of the RXs of the BS node $b$ in baseband. The digital canceller $\mathbf{D}_b$ contains the reciprocal residual SI channel and therefore, based on the considered model in \eqref{Eq: sig1}, is given by $\mathbf{D}_b= -\left(\mathbf{H}_{b,b}+\mathbf{C}_{b}\right)$. 

After applying A/D SI cancellation at BS's reception side, the received baseband training signals are used for estimating the UL channel. It is noted that, for the case where ${s}_u$ refers to information data symbol, BS node $b$ deploys the RX combiner $\mathbf{w}_{b}\in \mathbb{C}^{N_b\times 1}$ to estimate it as $\Hat{s}_u \triangleq \mathbf{w}_{b}^{\rm T}\mathbf{y}_{b}$. Similar to BS, the induced SI signal at the RX of the UE node is suppressed using A/D cancellation. Particularly, symbol $\Hat{s}_b$, which refers to the estimation for ${s}_b$, is computed in baseband as
\begin{equation}\label{Eq: sig2}
    \begin{split}
        \Hat{s}_{b} \triangleq \left({h}_{u,u}+{c}_{u}+{d}_u\right){s}_{u} + \mathbf{h}_{u,b}\mathbf{v}_b {s}_{b} + {n}_{u}, 
    \end{split}
\end{equation}
with ${n}_{u}$ denoting the zero-mean AWGN with variance $\sigma_{u}^2$. The single-tap analog canceller ${c}_{u}$ is designed as ${c}_{u}= -{h}_{u,u}$ \cite{duarte2012experiment}. Similar to the BS node $b$, the digital canceller at the UE node $u$ is mathematically expressed as ${d}_u = -({h}_{u,u}+{c}_u)$.

In practice, the precoding vector $\mathbf{v}_b$ is designed based on the estimated DL channel. In this paper, we capitalize on the UL/DL channel reciprocity and perform CSI estimation with training signals for the UL channel. The latter estimation is used for deriving $\mathbf{v}_b$. For the SI channels at both the BS and UE, we assume channel estimation when there are no active UL and DL transmissions; the latter channels are expected to changed rather slowly. In the following subsection, we present a CSI estimation error model for the UL training scheme and derive the average achievable DL rate.

\section{Joint Data Communication and CSI Estimation}\label{Subsec: FD_1}
\begin{figure}[!t]
\centering
\includegraphics[width=0.43\textwidth]{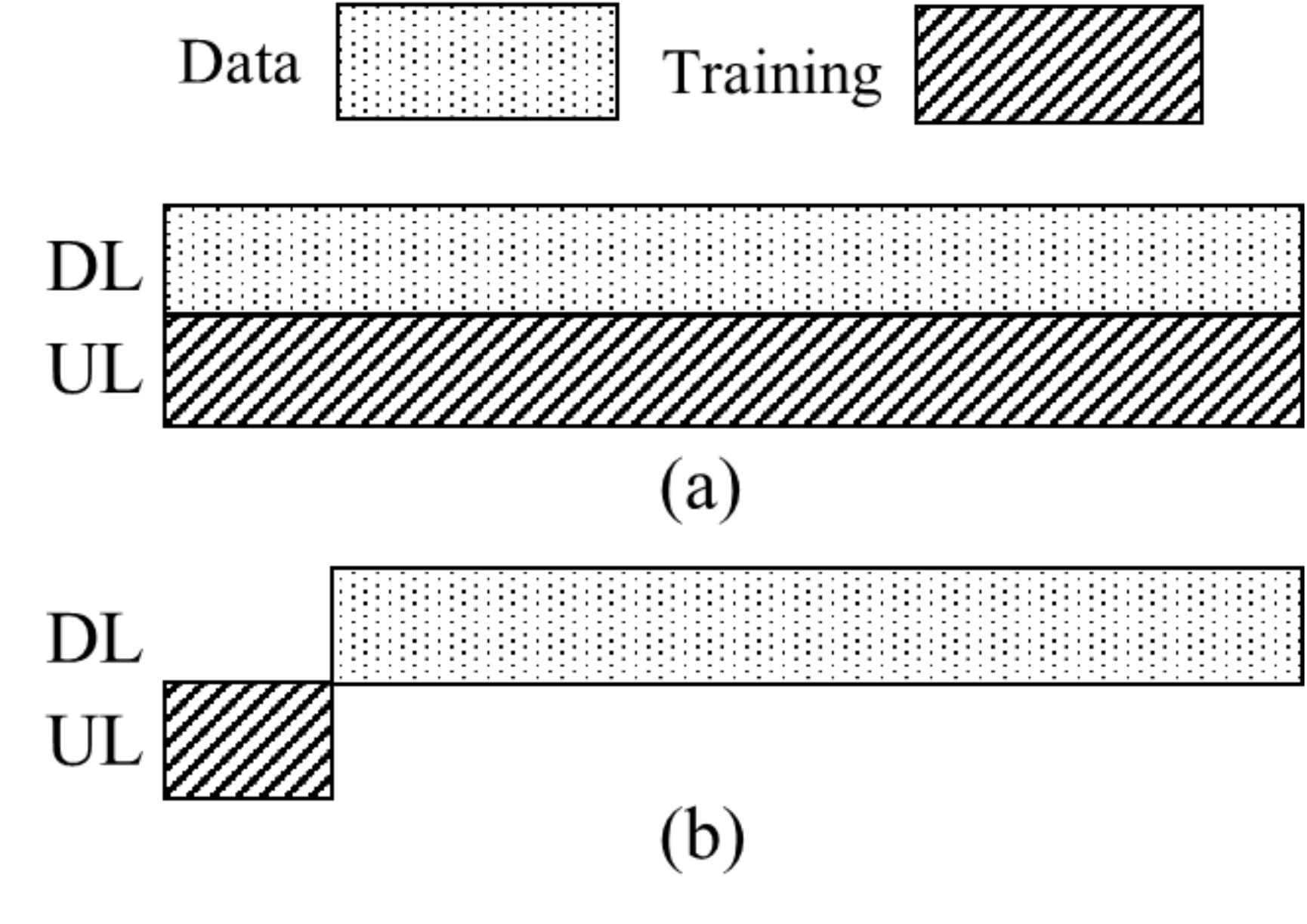}
\caption{(a) FD-enabled simultaneous DL information data and UL training data transmissions. (b) Conventional HD-based DL and UL transmissions.}
\label{fig: transmission_scheme}
\end{figure}

The considered FD operation at both BS node $b$ and UE node $u$ enables concurrent transmissions as well as concurrent receptions given sufficient SI cancellation. In this paper, we focus on the transmission scheme illustrated in Fig.~\ref{fig: transmission_scheme}(a), where $T$ information data symbols are transmitted in the DL simultaneously with $T$ training symbols communicated in the UL direction. As seen from this figure, the whole UL packet spanning $T$ symbols is dedicated for CSI estimation leading naturally to increased estimation accuracy. Alternatively, portion of the UL packet could be used for UL data transmission; this is left for future work. Exploiting UL/DL channel reciprocity, the CSI estimation is used for designing the DL precoding vector.  

Let the $T$-element column vectors $\mathbf{s}_{b,d}$ and $\mathbf{s}_{u,t}$ include respectively the information data symbols intended for DL communication and the training symbols used for UL CSI estimation. Based on \eqref{Eq: sig1}, the received training symbols at BS node $b$ can be grouped in $\mathbf{Y}_{b,t}\in \mathbb{C}^{N_b\times T}$, each on $\mathbf{Y}_{b,t}$'s column. For proper FD-based reception \cite{alexandropoulos2017joint, islam2019unified}, the RXs' RF chains need to be unsaturated from any residual SI stemming out analog SI cancellation. This fact translates to residual SI power constraints at the RXs of both considered nodes. We denote by $\lambda_b$ and $\lambda_u$ the residual SI power thresholds after analog cancellation and impose the specific constraints ${P}_{b}|[\left(\mathbf{H}_{b,b}+\mathbf{C}_{b}\right)\mathbf{v}_b]_{(i,:)}|^2\leq \lambda_b$ $\forall$$i=1,2,\dots,N_b$ RX of BS node $b$, and ${P}_u|\left({h}_{b,b}+{c}_{u}\right)|^2\leq \lambda_u$ at the RX of UE node $u$. When the latter constraints are met, SI has been properly degraded enabling successful reception.

\subsection{Proposed FD-Based Transmission Scheme}
Assuming that all $N_b$ RX RF chains at BS node $b$ are not saturated from SI, the Minimum Mean Square Error (MMSE) estimate for the UL channel ${\mathbf{h}}_{b,u}$ is derived as
\begin{equation}\label{Eq: est_real_chan}
    \begin{split}
        \widehat{\mathbf{h}}_{b,u}=(1+\mathbf{s}_{b,t}^{\rm H}\mathbf{s}_{b,t})^{-1}\mathbf{Y}_{b,t}\mathbf{s}_{b,t}^{*}.
    \end{split}
\end{equation}
Using UL/DL channel reciprocity, the DL channel estimation can be then obtained as $\widehat{\mathbf{h}}_{u,b}=\widehat{\mathbf{h}}_{b,u}^{\rm T}$. It can be shown that the actual DL channel and its latter estimation relate as \cite{iimori2019mimo,jindal2006mimo}   
\begin{equation}\label{Eq: actual_chan_relation}
    \begin{split}
        \mathbf{h}_{u,b} = \sqrt{1-\tau_{\text{UL}}^2}\widehat{\mathbf{h}}_{u,b} + \tau_{\text{UL}}{\mathbf{e}}_{u,b},
    \end{split}
\end{equation}
where ${\mathbf{e}}_{u,b}\in\mathbb{C}^{1\times N_b}$ is the estimation error vector having IID elements each modeled as $\mathcal{C}\mathcal{N}(0,l_{u,b})$, and $\tau_{\text{UL}}\in[0,1]$ is the Gauss-Markov error parameter that depends on the effective UL Signal-to-Noise Ratio (SNR). The case $\tau_{\rm UL}=0$ implies ideal DL CSI, whereas $\tau_{\rm UL} = 1$ signifies unavailable channel estimation. The Mean Squared Error (MSE) of DL CSI estimation at BS node $b$ is given by
\begin{equation}\label{Eq: sig5}
    \begin{split}
        {\rm MSE_{FD}}\triangleq\tau_{\text{UL}}^2=\left(1+T\frac{{P}_u \|{\mathbf{h}}_{u,b}\|^2}{\sigma^2_b + \sigma^2_{r,b}}\right)^{-1},
    \end{split}
\end{equation}
where $\sigma^2_{r,b}\triangleq {P}_{b}\|\left(\widehat{\mathbf{H}}_{b,b}+\mathbf{C}_{b}+\mathbf{D}_b\right)\mathbf{v}_b \|^2$ is the residual SI signal power after both A/D SI cancellation.

Using the estimation $\widehat{\mathbf{h}}_{u,b}$, the achievable DL rate for the considered FD-based transmission scheme is calculated as
\begin{equation}\label{Eq: achievable_rate_FD_1}
    \begin{split}
        \mathcal{R}_{\text{DL}}^{(\text{FD})}\triangleq \log_2\left(1+ \frac{(1-\tau_{\text{UL}}^2){P}_{b}|\widehat{\mathbf{h}}_{u,b}\mathbf{v}_b|^2}{\sigma^2_{u}+ \sigma^2_{r,u} +\tau_{\text{UL}}^2 {P}_{b}l_{u,b}\|\mathbf{v}_b\|^2 }\right)
    \end{split}
\end{equation}
with $\sigma^2_{r,u}\triangleq{P}_u|\left({h}_{u,u}+{c}_{u}+{d}_u\right)|^2$ being the residual SI signal power after both A/D SI cancellation at UE node $u$.

\subsection{HD-Based Downlink and Uplink Transmissions}\label{subsec: HD}
For comparison purposes, we consider the HD-based transmission scheme of Fig.~\ref{fig: transmission_scheme}(b), where part of a packet duration is used for UL CSI estimation and the rest of DL data communication. We specifically assume that 
$T_{\rm HD}$ symbols out of the total $T$ spanning a packet are dedicated for UL channel sounding. The remaining $T-T_{\rm HD}$ symbols are devoted for DL data transmission from BS node $b$ to UE node $u$. Similar to \eqref{Eq: est_real_chan}, the MMSE estimate of $\mathbf{h}_{b,u}$ is obtained using $T_{\rm HD}$ training symbols, and relying on the UL/DL channel reciprocity, the actual DL channel and its estimation are related as
\begin{equation}\label{Eq: est_hd_chan}
    \begin{split}
        \mathbf{h}_{u,b} = \sqrt{1-\left(\tau_{\text{UL}}^{({\rm HD})}\right)^2}\,\widehat{\mathbf{h}}_{u,b} + \tau_{\text{UL}}^{({\rm HD})}{\mathbf{e}}_{u,b},
    \end{split}
\end{equation}
where $\tau_{\text{UL}}^{({\rm HD})}$ is the error parameter for UL CSI estimation. Using the latter expression, the MSE of DL CSI estimation with this HD-based transmission scheme is given by
\begin{equation}\label{Eq: error_HD}
    \begin{split}
        {\rm MSE_{HD}}\triangleq\left(\tau_{\text{UL}}^{({\rm HD})}\right)^2=\left(1+T_{\rm HD}\frac{{P}_u \|{\mathbf{h}}_{b,u}\|^2}{\sigma^2_b}\right)^{-1},
    \end{split}
\end{equation}
whereas the achievable DL rate is obtained as
\begin{equation}\label{Eq: achievable_rate_HD}
    \begin{split}
        \mathcal{R}_{\text{DL}}^{(\text{HD})}\triangleq \log_2\left(1+ \frac{(1-\left(\tau_{\text{UL}}^{({\rm HD})}\right)^2){P}_{b}|\widehat{\mathbf{h}}_{u,b}\mathbf{v}_b|^2}{\sigma^2_{u} +\left(\tau_{\text{UL}}^{({\rm HD})}\right)^2 {P}_{b}l_{u,b}\|\mathbf{v}_b\|^2 }\right).
    \end{split}
\end{equation}

\section{Proposed Joint Optimization Framework}
\begin{algorithm}[!t]
    \caption{Proposed FD MIMO Design}
    \label{alg:the_alg}
    \begin{algorithmic}[1]
        \renewcommand{\algorithmicrequire}{\textbf{Input:}}
       \renewcommand{\algorithmicensure}{\textbf{Output:}}
        \REQUIRE $\widehat{\mathbf{H}}_{b,b}$, $\widehat{\mathbf{h}}_{u,b}$, $\widehat{{h}}_{u,u}$, ${P}_b$, ${P}_u$, $N$, and $N_b$.
        \ENSURE $\mathbf{v}_b$, $\mathbf{C}_b$, $c_u$,  $\mathbf{D}_b$, and $d_u$.
        \STATE Obtain the $N$-tap analog canceller $\mathbf{C}_b$ using \eqref{Eq: analog_canc}.
        \STATE Obtain $\mathbf{Q}_b$ including the $N_b$ right-singular vectors of $(\widehat{\mathbf{H}}_{b,b} + \mathbf{C}_b)$ corresponding to the singular values in descending order.
        \STATE Set ${c}_{u}= -\widehat{h}_{u,u}$.
        \FOR{$\alpha=N_b, N_b-1, \dots, 2$}
            \STATE Set $\mathbf{F}_b=[\mathbf{Q}_b]_{(:,N_b-\alpha+1 : N_b)}$.
            \STATE Set $\mathbf{g}_b = \frac{\mathbf{F}_b^T\widehat{\mathbf{h}}_{u,b} }{\|\mathbf{F}_b^T\widehat{\mathbf{h}}_{u,b}\|}$.
            \STATE Set the DL precoder as $\mathbf{v}_b= \mathbf{F}_b \mathbf{g}_b$.
            \IF {${P}_{b}|[(\widehat{\mathbf{H}}_{b,b}+\mathbf{C}_{b})\mathbf{v}_b]_{(i,:)}|^2\leq \lambda_b, \forall_{i=1,\dots,N_b}$ and ${P}_u|(\widehat{h}_{u,u}+{c}_{u})|^2\leq \lambda_u$} 
            \STATE Output $\mathbf{v}_b$, $\mathbf{C}_b$, $c_u$, $\mathbf{D}_b=-(\widehat{\mathbf{H}}_{b,b}+ \mathbf{C}_b)$, $d_{u}=-(\widehat{{h}}_{u,u}+c_{u})$, and stop the algorithm.
            \ENDIF
        \ENDFOR
        \STATE Set $\mathbf{v}_b=[\mathbf{Q}_b]_{(:,N_b)}$.
        \IF {${P}_{b}|[(\widehat{\mathbf{H}}_{b,b}+\mathbf{C}_{b})\mathbf{v}_b]_{(i,:)}|^2\leq \lambda_b, \forall_{i=1,\dots,N_b}$ and ${P}_u|(\widehat{h}_{u,u}+{c}_{u})|^2\leq \lambda_u$} 
            \STATE Output $\mathbf{v}_b$, $\mathbf{C}_b$, $c_u$, $\mathbf{D}_b=-(\widehat{\mathbf{H}}_{b,b}+ \mathbf{C}_b)$, $d_{u}=-(\widehat{{h}}_{u,u}+c_{u})$, and stop the algorithm.
        \ELSE
        \STATE Output that the $\mathbf{C}_b$ realizations or $c_{u}$ do not meet the receive RF saturation constraints.
        \ENDIF
    \end{algorithmic}
\end{algorithm}
We focus on the joint design of the digital TX precoder $\mathbf{v}_b$, the analog SI cancellers $\mathbf{C}_b$ and ${c}_u$, as well as the digital cancellers $\mathbf{D}_b$ and ${d}_{u}$ at both BS node $b$ and UE node $u$ maximizing the estimated achievable DL rate. In mathematical terms, the considered optimization problem is expressed as: 
\begin{align}\label{eq: optimization_eq}
        \nonumber\underset{\substack{\mathbf{v}_b,\mathbf{C}_b, {c}_u\\ \mathbf{D}_b, {d}_{u}}}{\text{max}} &\log_2\left(1+ \frac{(1-\tau_{\text{UL}}^2){P}_{b}|\widehat{\mathbf{h}}_{u,b}\mathbf{v}_b|^2}{\sigma^2_{u}+ \sigma^2_{r,u} +\tau_{\text{UL}}^2 {P}_{b}l_{u,b}\|\mathbf{v}_b\|^2 }\right)\\
        \text{\text{s}.\text{t}.}\quad
        &P_b|[(\widehat{\mathbf{H}}_{b,b}+\mathbf{C}_{b})\mathbf{v}_b ]_{(i,:)}|^2\leq \lambda_b \,\forall i=1,\dots,N_b,\\
        &P_u|\left(\widehat{{h}}_{u,u}+{c}_{u}\right)|^2\leq \lambda_u, \nonumber\\
        &\mathbb{E}\{|\mathbf{v}_b s_{b}|^2\}\leq {P}_{b},\,\,\text{and}\,\,\,
        \mathbb{E}\{|s_{u}|^2\}\leq {P}_{u}.\nonumber
\end{align}
In this formulation, the first constraint imposes the analog SI cancellation saturation threshold $\lambda_{b}$ at BS node $b$. As previously discussed, this threshold ensures proper reception of the training symbols by all $N_b$ RX RF chains of the BS, which means that the UL channel can be efficiently estimated using \eqref{Eq: est_real_chan}. The second constraint enforces the saturation threshold $\lambda_{u}$ at UE node $u$ assuring feasible decoding of BS's information data symbols. The final two constraints in \eqref{eq: optimization_eq} refer to the nodes' average transmit powers.

The optimization problem in \eqref{eq: optimization_eq} is quite difficult to tackle, since it is non-convex including couplings among the optimization variables. We propose in this paper an alternating optimization approach to solve it suboptimally, leaving other possibilities for future work. To this end, we start with an allowable $\mathbf{C}_{b}$ realization given the available number of analog canceller taps $N$, where the tap values are set to be the respective amplitude elements of the estimated SI channel $\widehat{\mathbf{H}}_{b,b}$. Based on the chosen $\mathbf{C}_{b}$, we seek for the precoding vector $\mathbf{v}_b$ maximizing the DL rate, while meeting the first constraint for the BS analog SI cancellation threshold $\lambda_{b}$. This procedure is repeated for all allowable realizations of $\mathbf{C}_{b}$ to find the best pair of $\mathbf{C}_{b}$ and $\mathbf{v}_b$. Adopting the approach in \cite{alexandropoulos2017joint}, the BS precoder for DL data communication is constructed as $\mathbf{v}_b= \mathbf{F}_b \mathbf{g}_b$, where $\mathbf{F}_b \in \mathbb{C}^{N_b \times \alpha}$ aims at abating the residual SI after analog cancellation and $\mathbf{g}_b \in \mathbb{C}^{\alpha \times 1}$ is the optimum precoder maximizing the rate of the effective DL channel $\widehat{\mathbf{h}}_{u,b}\mathbf{F}_b$. The parameter $\alpha$ is a positive integer taking the values $1 \leq \alpha \leq N_b$. At the UE node $u$, a single-tap analog SI canceller is implemented by designing its tap value as ${c}_{u}= -\widehat{h}_{u,u}$. To maximize the signal-to-interference-plus-noise ratio, the residual SI is further reduced by setting the digital cancellation signal at both FD nodes as their respective complementary residual SI channels after analog SI cancellation. The proposed solution for the considered optimization problem \eqref{eq: optimization_eq} is summarized in Algorithm \ref{alg:the_alg}.

\section{Simulation Results and Discussion}
In this section, we investigate the performance of the proposed FD-enabled simultaneous DL information data and UL training
data transmission scheme discussed Sec.~\ref{sec: Sig1}. The simulation parameters and assumptions are detailed in the following Sec.~\ref{subsec: sim_param}, while Sec.~\ref{subsec: achievable_rate} presents representative results on the achievable DL rate and UL channel estimation.
\begin{figure}[!tpb]
\centering
\includegraphics[width=0.46\textwidth]{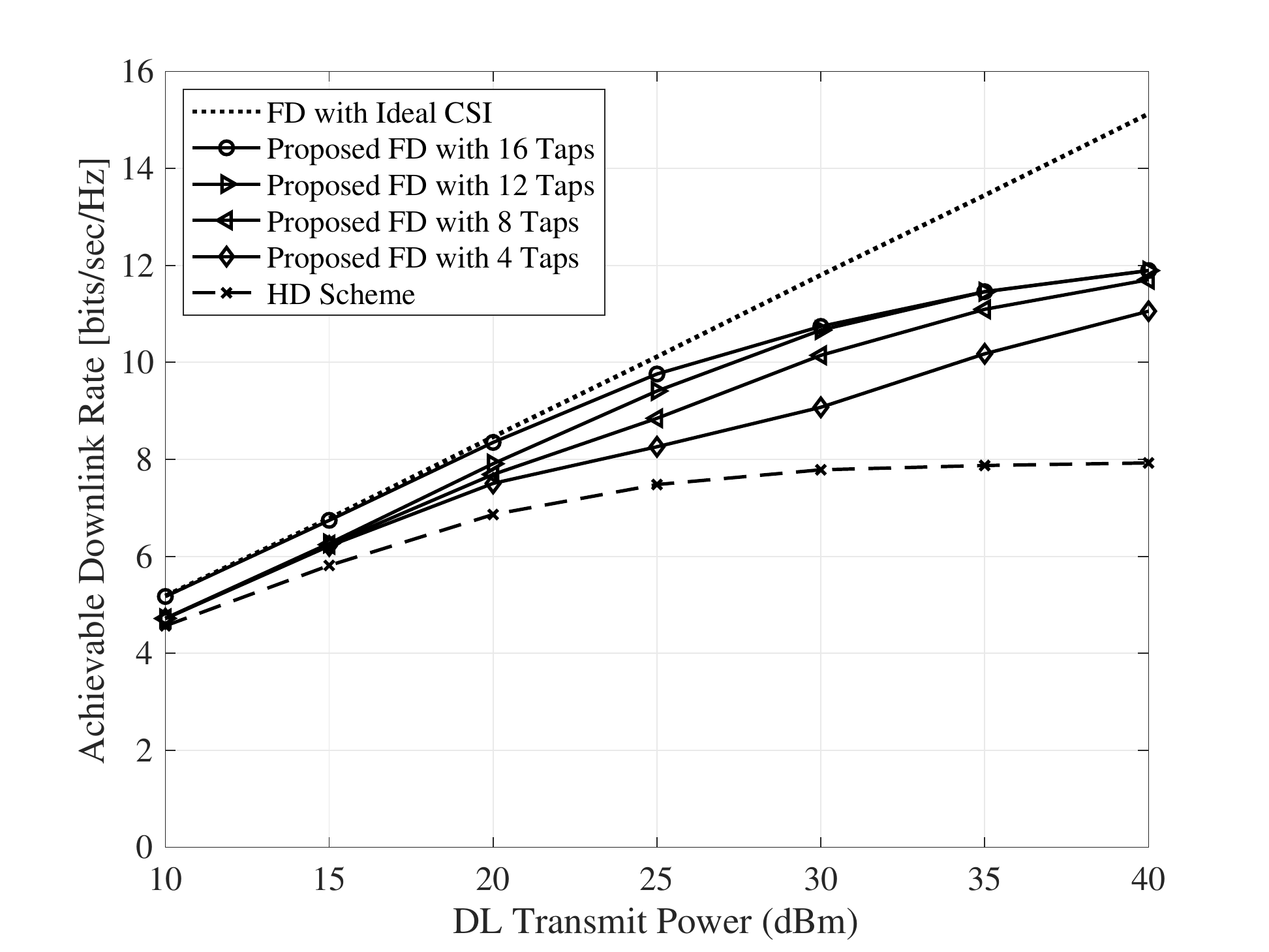}
\caption{Achievable DL rate versus the DL transmit power in dBm for the case of limited UL transmit power at $5$dBm.}
\label{fig: rate_vs_DL_power}
\end{figure}

\subsection{Simulation Parameters}\label{subsec: sim_param}
We perform an extensive simulation following the FD MIMO architecture illustrated in Fig$.$~\ref{fig: transmission}. We have considered a $4\times4$ (i.e., $N_b=4$) FD MIMO BS node $b$ serving a FD single-antenna UE node $u$. Both DL and UL channels are assumed as block Rayleigh fading channels with a pathloss of $110$dB. The SI channels at both FD nodes $b$ and $u$ are simulated as Rician fading channel with a $K$-factor of $35$dB and pathloss of $40$dB \cite{duarte2012experiment}. We have considered a narrowband communication system with a bandwidth of $1.4$MHz, which is a supported bandwidth for Long Term Evolution (LTE). RX noise floors at both nodes were assumed to be $-110$dBm. To this end, the RXs have effective dynamic range of $62.24$dB provided by the $14$-bit analog-to-digital converters (ADC) for a Peak-to-Average-Power-Ratio (PAPR) of $10$ dB \cite{AD3241}. Therefore, the residual SI power after analog SI cancellation at the input of each RX chain has to be below $-47.76$dBm to avoid saturation. Furthermore, non-ideal multi-tap analog canceller is considered with steps of $0.02$dB for attenuation and $0.13^\circ$ for phase as in \cite{alexandropoulos2017joint}. We have used $1000$ independent Monte Carlo simulation runs to calculate the performance of all considered designs. In each run, a total number of $T=400$ symbols were considered in every transmission packet. For the compared HD scheme, $10\%$ the packet's symbols (i.e., $T_{\rm HD}=40$) were dedicated for UL channel sounding.

\begin{figure}[!tpb]
\centering
\includegraphics[width=0.46\textwidth]{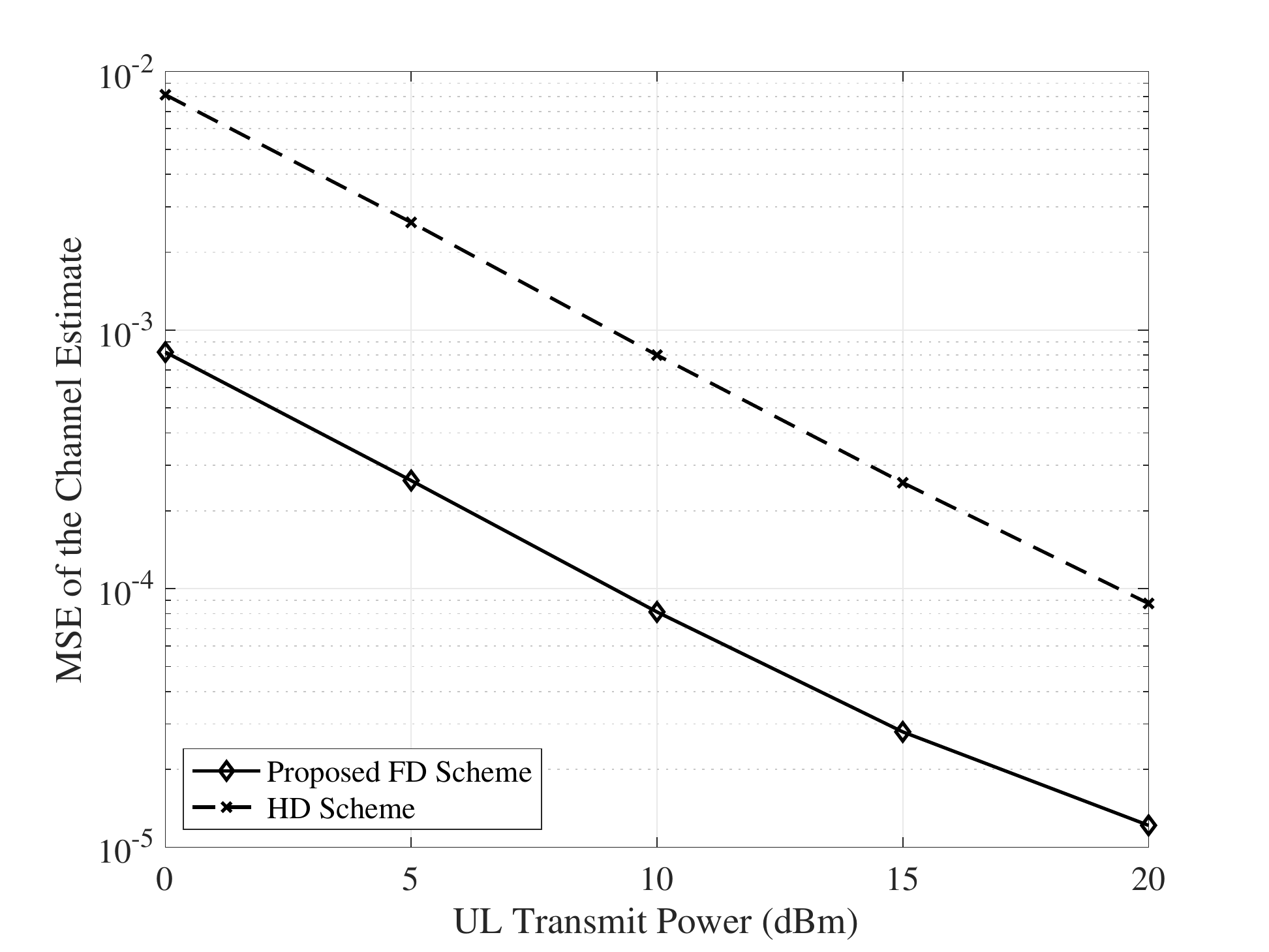}
\caption{MSE of UL channel estimation versus the UL transmit power for the case where the DL transmit power is $40$dBm.}
\label{fig: rate_vs_analog_taps}
\end{figure}

\subsection{Achievable DL Rate and CSI Estimation Performances}\label{subsec: achievable_rate}
Figure~\ref{fig: rate_vs_DL_power} depicts the achievable DL rate performance as a function of the DL transmit power in dBm for the case where the UL transmit power for training data is $5$dBm with $50$dB digital SI cancellation. We have considered different numbers for the analog canceller taps $N$ for the proposed FD MIMO architecture at BS node $b$, and also sketched the performance of the ideal CSI case with $N=16$ together with the HD-based transmission scheme discussed in Sec.~\ref{Subsec: FD_1}. It is shown in the figure that, for low to moderate transmit powers ($\leq20$dBm), the proposed FD-based transmission scheme exhibits similar performance to the ideal CSI case. The performance gap in large transmit powers is a consequence of practical CSI estimation error and residual SI at the FD RXs. For the highest considered transmit power of $40$dBm, Algorithm~1 results in less than $1$bit/sec/Hz DL rate gap between $4$ and $16$ taps for analog SI cancellation, leading to $75\%$ reduction in the hardware complexity for the analog canceller. It is also evident that, for all DL transmit powers, the proposed scheme outperforms HD-based transmissions. For example, when the DL transmit power is $40$dBm, the DL data rate with FD-based transmissions is 1.4$\times$ the HD-based one. 

\begin{figure}[!tpb]
\centering
\includegraphics[width=0.46\textwidth]{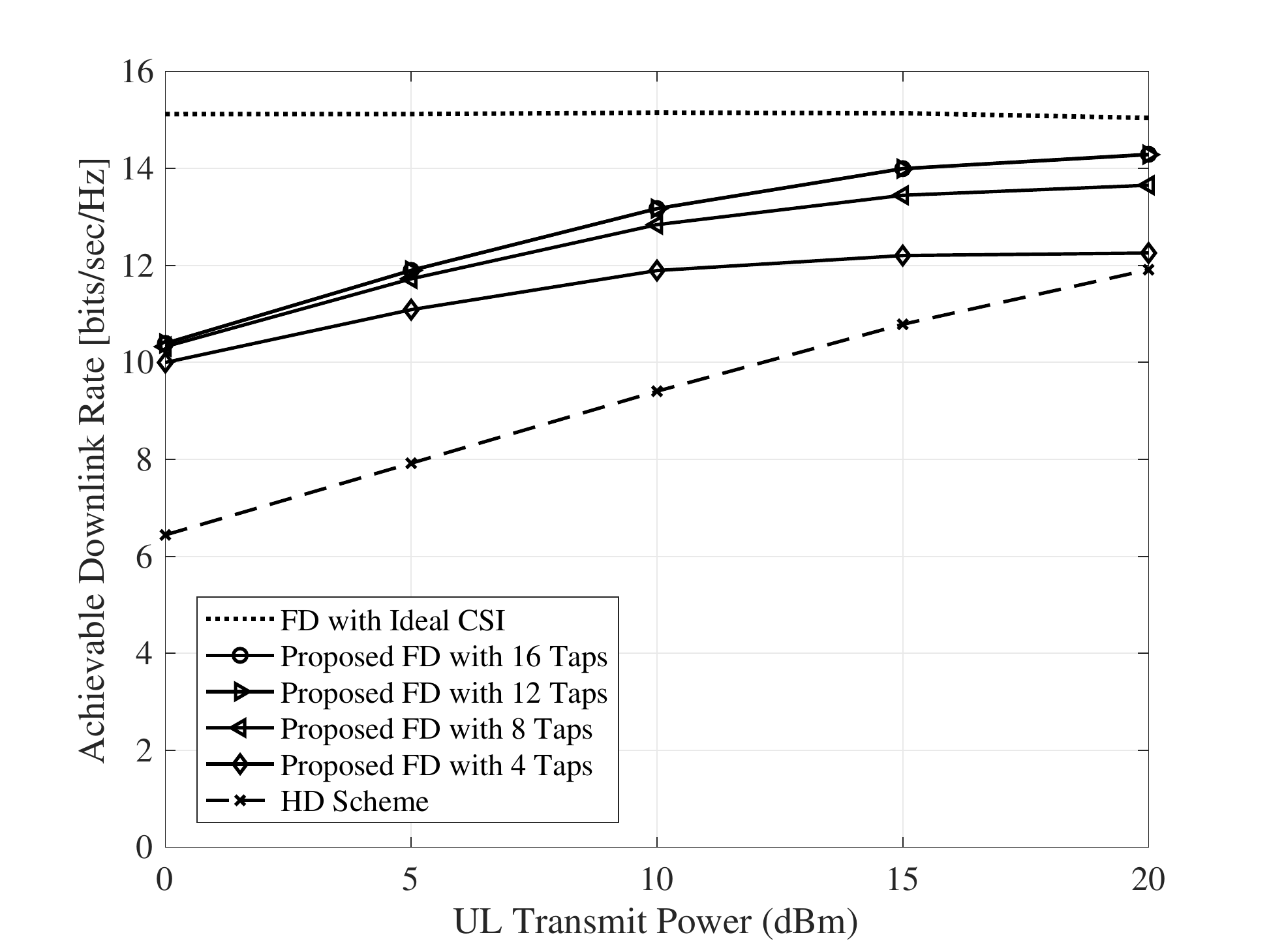}
\caption{Achievable DL rate versus the UL transmit power for the case where the DL transmit power is $40$dBm.}
\label{fig: rate_vs_UL_power}
\end{figure}

The MSE of CSI estimation versus the UL transmit power in dBm is shown in Fig.~\ref{fig: rate_vs_analog_taps} for the case where the DL transmit power is $40$dBm. It is noted that the proposed FD-based scheme provides similar MSE performance irrespective of $N$, since it targets reducing SI at BS's RX below the noise floor for all cases. Despite the residual SI, the proposed scheme exhibits substantially smaller MSE compared to the HD scheme, since the former supports simultaneous communication of $400$ ($T=400\gg T_{\rm HD}=40$) UL training symbols together with $400$ DL information data symbols. Finally, Fig.~\ref{fig: rate_vs_UL_power} plots the achievable DL rate for different UL transmit powers and DL transmit power of $40$dBm. It is shown that with increasing UL transmit power, the DL rate of the proposed FD-based scheme with CSI estimation approaches that of the ideal CSI case, which witnesses that the CSI error decreases. The same trend happens for the HD-based transmission scheme, whose DL rate performance is however outperformed from the proposed scheme with only $4$ analog SI cancellation taps. For UL transmit powers larger than $12.5$dB, the DL rate performance of the proposed scheme is around $1$bit/sec/Hz close to the ideal CSI case with 50$\%$ reduction in the number of the analog taps. 

\squeezeup
\section{Conclusion and Future Work}
In this paper, we proposed a point-to-point two-way FD MIMO communication system for simultaneous DL information data transmission and UL CSI estimation with reduced complexity multi-tap analog SI cancellation. Considering a MMSE-based channel estimation error model, we presented a unified optimization framework for the joint design of digital TX precoding and A/D SI cancellation. Our performance evaluation results demonstrated that the proposed transmission protocol is capable of achieving improved achievable DL rates compared to HD systems with reduced complexity analog cancellation compared to conventional FD MIMO architectures. For future work, we intend to extend our FD-based transmission scheme to multi-user systems considering also UL information data communication.
\section*{Acknowledgments}
This work was partially funded by the National Science Foundation CAREER award \#1620902.

\squeezeupann

\bibliographystyle{IEEEtran}
\bibliography{IEEEabrv,ms}
\end{document}